\documentclass[a4paper, amsfonts, amssymb, amsmath, reprint, showkeys, nofootinbib, twoside, onecolumn,notitlepage]{revtex4-1}

\def\apj{{ApJ}}                 

\def\prd{{Phys.~Rev.~D}}        
\def\prl{{Phys.~Rev.~Lett.}}    
\def\jcp{{J.~Chem.~Phys.}}      

\usepackage{amsmath,amstext}
\usepackage[T1]{fontenc}
\usepackage{amssymb}
\input{epsf}
\usepackage{graphicx}
\usepackage{ae,aecompl}

\usepackage{hyperref}
\usepackage{amsmath}
\usepackage{amssymb}
\usepackage{mathtools}
\usepackage{bm}
\usepackage{cleveref}
\usepackage{tensor}
\usepackage{braket}
\usepackage{enumitem}
\usepackage{mhchem}
\usepackage{cancel}

\usepackage{color}

\newcommand{\spc}{\quad \quad \quad}

\def\be{\begin{equation}}
\def\ee{\end{equation}}
\def\beq{\begin{eqnarray}}
\def\eeq{\end{eqnarray}}

\begin{document}
\title{Relativistic bulk viscous fluids of Burgers type and their presence in neutron stars}
\author{L.~Gavassino}
\affiliation{
Department of Mathematics, Vanderbilt University, Nashville, TN, USA
}

\begin{abstract}
It is well known that a mixture of two chemical components undergoing one chemical reaction is a bulk viscous fluid, where the bulk stress evolves according to the Israel-Stewart theory. Here, we show that a mixture of three independent chemical components undergoing two distinct chemical reactions can also be viewed as a bulk viscous fluid, whose bulk stress now is governed by a second-order differential equation which reproduces the Burgers model for viscoelasticity. This is a rigorous and physically motivated example of a fluid model where the viscous stress does not undergo simple Maxwell-Cattaneo relaxation, and can actually overshoot the Navier-Stokes stress. We show that, if one accounts for muons,  neutron star matter is indeed a bulk viscous fluid of Burgers type.
\end{abstract}

\maketitle

\section{Introduction}

As a branch of fluid mechanics, ``rheology'' is the study the evolution of the stress tensor outside of the Navier-Stokes regime \cite{Malkin_book}. To understand what this means in practice, consider the example of the bulk viscous stress $\Pi$. If   a fluid element undergoes a small time-dependent expansion, with expansion rate $\nabla_\mu u^\mu (t)$, where $t$ is the proper time along the element's worldline, then we should be able to write (in a regime of linear response) an equation of the form \cite{Denicol_Relaxation_2011}
\begin{equation}\label{babibbo}
\Pi(t)= -\int_{-\infty}^{+\infty} \! \! \! G(t')\nabla_\mu u^\mu(t-t') dt' \, ,
\end{equation}
where the kernel $G(t')$ is a Green-function. Now, the fluid is said to be ``in the Navier-Stokes regime'' \cite{LindblomRelaxation1996} when the expansion is so slow that we can approximate $\nabla_\mu u^\mu(t-t') \approx \nabla_\mu u^\mu(t)$ over the support of $G(t')$, so that we have 
\begin{equation}\label{Pippo}
\Pi(t)\approx - \zeta \, \nabla_\mu u^\mu(t) \, ,
\end{equation}
where $\zeta=\int_{-\infty}^{+\infty}  G(t') dt'$ is the bulk viscosity coefficient \cite{landau6}. As we can see, in the Navier-Stokes regime all fluids behave in a similar way \cite{Geroch1995}, and the complexity of the function $G(t')$ is fully reabsorbed in the transport coefficient $\zeta$. However, outside of this infinitely slow regime, the details of $G(t')$ become important, and different fluids may behave in very different ways, giving rise to a plethora of different possible mechanical models \cite{Geroch_Lindblom_1991_causal,GavassinoNonHydro2022}. The main goal of rheology is to classify all fluids based on the detailed structure of their Green function \cite{Malkin_book}.

The simplest example of a rheological model is the Israel-Stewart theory \cite{Israel_Stewart_1979,Hishcock1983,Causality_bulk}, which in non-relativistic rheology is called ``Maxwell model'' \cite{maxwell_1867,Findley_book,BaggioliHolography}. It posits that the bulk stress obeys a relaxation-type equation of motion of the form $\tau u^\mu \nabla_\mu \Pi +\Pi = -\zeta \nabla_\mu u^\mu$ (where $\tau$ is a relaxation time), which corresponds to choosing the following Green function:
\begin{equation}\label{ISGreen}
G(t')=\dfrac{\zeta}{\tau}\Theta(t')e^{-t'/\tau} \, . 
\end{equation}
There are indeed many fluids that can be rigorously described (in some regimes of interest) by the Israel-Stewart theory for bulk viscosity, e.g. two-temperature systems \cite{GavassinoFronntiers2021}, diatomic gases \cite{Tisza_Bulk}, superfluids \cite{GavassinoKhalatnikov2022}, and fluid mixtures undergoing one single chemical reaction \cite{BulkGavassino}. However, being a rheological model (i.e., arising from a specific choice of $G$), the Israel-Stewart theory cannot share the same universal character as the Navier-Stokes equation \eqref{Pippo}. In fact, in general there is no universal ``Israel-Stewart regime''\footnote{A commonly quoted derivation of the Israel-Stewart theory \cite{Romatschke2017} consists of expanding $\nabla_\mu u^\mu(t{-}t')$ to linear order in $t'$, namely $\nabla_\mu u^\mu(t{-}t')\approx \nabla_\mu u^\mu(t)-t' u^\alpha \nabla_\alpha [\nabla_\mu u^\mu(t)]$, so that equation \eqref{babibbo} becomes $\Pi \approx -\zeta \nabla_\mu u^\mu -b u^\alpha \nabla_\alpha (\nabla_\mu u^\mu)$, with $b=-\int_{-\infty}^{+\infty} G(t')t'dt'$. Then, one invokes equation \eqref{Pippo} to approximate the second term: $\Pi \approx -\zeta \nabla_\mu u^\mu +b\zeta^{-1} u^\alpha \nabla_\alpha \Pi$. However, it is evident that such derivation works only in the Navier-Stokes regime (infinitely slow processes), and it cannot be invoked to justify the ``universality'' of the Israel-Stewart theory in a rheological context. Indeed, from a rheological perspective, the model $\Pi \approx -\zeta \nabla_\mu u^\mu -b u^\alpha \nabla_\alpha (\nabla_\mu u^\mu)$ is profoundly different from the Israel-Stewart theory, since its Green function is $G(t')=\zeta\delta(t')-b \, \delta'(t')$, which differs from \eqref{ISGreen}.} \cite{Geroch2001}. 
A fluid is either of Israel-Stewart type, or not, depending on the dynamics of its non-equilibrium degrees of freedom \cite{GavassinoUniversalityI2023,GavassinoUniversalityII2023}. For example, a well-known alternative to the Maxwell model is the Burgers model \cite{Findley_book}, whose Green function is the sum of two Israel-Stewart Green functions \cite{Malek2018}:
\begin{equation}\label{GtB}
G(t')= \dfrac{\zeta_{(1)}}{\tau_{(1)}}\Theta(t')e^{-t'/\tau_{(1)}} + \dfrac{\zeta_{(2)}}{\tau_{(2)}}\Theta(t')e^{-t'/\tau_{(2)}} \, . 
\end{equation}

The goal of this article is to provide a rigorous example of a relativistic bulk viscous fluid of Burgers type. We will use the mathematical correspondence between chemistry and bulk viscosity \cite{BulkGavassino} to show that a fluid mixture with two non-conserved chemical affinities is ``dual'' to the Burgers rheological model for bulk viscosity. As a consequence, a three-component fluid mixture of this kind cannot be described by the Israel-Stewart theory outside of the Navier-Stokes regime (even close to equilibrium), but it requires the use of a different equation of motion for $\Pi$, which is of second order in time. As a quick application, we shall also show that neutron star matter is indeed governed by Burgers-type bulk viscous dynamics, if we account for the presence of muons.

Throughout the article, we adopt the metric signature $(-,+,+,+)$, and work in natural units: $c=k_B=1$. 

\section{Reacting mixtures as bulk viscous fluids}\label{sec2}

It is well known \cite{landau6,Sawyer_Bulk1989,BulkGavassino} that chemical mixtures undergoing chemical reactions can be rigorously ``reinterpreted'' as bulk viscous fluids. In this section, we briefly review the proof, as given in \cite{BulkGavassino}.

\subsection{The dynamics of a mixture}\label{dynamo}


We consider a relativistic fluid mixture \cite{noto_rel} with a single conserved particle current $n^\mu$ (e.g. the baryon current) and an arbitrary number of non-conserved currents $n_a^\mu$, where $a$ is a chemical index. We assume that all the fluid tensors are isotropic in the rest frame, defined by the four-velocity $u^\mu \propto n^\mu$, so that the constitutive relations take the form \cite{MTW_book,noto_rel,Gourgoulhon2006,Landry2020}
\begin{equation}\label{tmunu}
\begin{split}
T^{\mu \nu} ={}& (\rho+P)u^\mu u^\nu + P g^{\mu \nu} \, , \\
s^\mu ={}& \mathfrak{s} n u^\mu \, , \\
n^\mu ={}& n u^\mu \, , \\
n_a^\mu ={}& Y_a n  u^\mu \, , \\
\end{split}
\end{equation}
where $T^{\mu \nu}$ is the stress-energy tensor and $s^\mu$ is the entropy four-current. The scalar fields $\rho$, $P$, $\mathfrak{s}$ and $Y_a$ are interpreted respectively as the energy density, the pressure, the specific entropy and the $a$-th non-conserved fraction (all measured in the rest frame). The thermodynamics of the fluid is most conveniently described in the ``per-particle representation''. In particular, defined the specific volume $v=n^{-1}$ and the specific energy $u=\rho/n$, we write an equation of state of the form $u=u(\mathfrak{s},v,Y_a)$, whose differential reads \cite{Callen_book} (we adopt Einstein's convention for repeated chemical indices)
\begin{equation}\label{duuu}
du = Td \mathfrak{s}-Pdv -\mathbb{A}^a dY_a \, ,
\end{equation} 
where $T$ is the temperature and $-\mathbb{A}^a$ is the chemical potential of the species $a$. Since the fractions $Y_a$ are not conserved, in chemical equilibrium we must have $\mathbb{A}^a=0$ (due to the maximum entropy principle \cite{landau5,Callen_book,GavassinoGibbs2021}), so that we can interpret $\mathbb{A}^a$ as reaction affinities \cite{PrigoginebookModernThermodynamics2014,peliti_book}. The equations of motion of the system are the conservation laws $\nabla_\mu T^{\mu \nu}=0$ and $\nabla_\mu n^\mu=0$, and the particle production equations: $\nabla_\mu n^\mu_a =\mathcal{R}_a$. If the fluid is not too far from local equilibrium, we can expand the reaction rates $\mathcal{R}_a$ to first order in the affinities, $\mathcal{R}_a=\Xi_{ab}\mathbb{A}^b$ \cite{carter1991,Termo}, so that we have the field equations
\begin{equation}\label{uY}
u^\mu \nabla_\mu Y_a = v \Xi_{ab}\mathbb{A}^b \, .
\end{equation}
Given that particle fractions are invariant under time reversal, the Onsager-Casimir principle \cite{Onsager_Casimir,GavassinoCasmir2022} requires that the reaction matrix $\Xi_{ab}$ be symmetric. Furthermore, it must also be non-negative definite, since the second law of thermodynamics ($\nabla_\mu s^\mu \geq 0$ \cite{Israel_2009_inbook}), combined with all other field equations, implies
\begin{equation}\label{us}
T u^\mu \nabla_\mu \mathfrak{s} = v\Xi_{ab}\mathbb{A}^a\mathbb{A}^b \geq 0 \, .
\end{equation} 
Since the fractions $Y_a$ are not conserved, we can actually assume that $\Xi_{ab}$ is positive definite, and therefore invertible. We call its matrix inverse $\Xi^{ab}$, so that $\Xi^{ab}\Xi_{bc}= \delta\indices{^a _c}$.

Now that we have stated the equations of motion of  relativistic fluid mixtures, we can proceed to prove that such substances are indeed bulk viscous fluids. In particular, we can show that, if the dynamics is sufficiently slow, solutions of the fluid equations asymptotically relax \cite{Geroch1995} to the constitutive relations of the relativistic Navier-Stokes theory \cite{Eckart40} with only bulk viscosity.

\subsection{Bulk viscous behaviour of mixtures}\label{dinamo2}

For the analysis that follows, it is particularly convenient to treat the collection of fields $\varphi_i =\{u^\mu, \mathfrak{s},v,\mathbb{A}^a \}$ as our independent degrees of freedom. This means that we will regard all the physical tensors in \eqref{tmunu} as functions of these fields. In particular, all thermodynamic quantities $f$ from now on are understood as functions $f(\mathfrak{s},v,\mathbb{A}^a)$, and partial derivatives are performed accordingly.  For example, if we write $\partial f/\partial v$, it is understood that the variables which are held constant are $\mathfrak{s}$ and $\mathbb{A}^a$. If we write $\partial f/\partial \mathbb{A}^a$, it is understood that we are holding constant $v$, $\mathfrak{s}$ and all other $\mathbb{A}^b$ (with $b \neq a$). Then, if the fluid is close to local thermodynamic equilibrium (i.e., if $\mathbb{A}^a$ are small), we can make the following first-order  expansions:
\begin{equation}
\begin{split}
\rho(\mathfrak{s},v,\mathbb{A}^a) \approx{}& \rho(\mathfrak{s},v,0) +\dfrac{\partial \rho}{\partial \mathbb{A}^a} \mathbb{A}^a \, , \\
P(\mathfrak{s},v,\mathbb{A}^a) \approx{}& P(\mathfrak{s},v,0) +\dfrac{\partial P}{\partial \mathbb{A}^a} \mathbb{A}^a \, .  \\ 
\end{split}
\end{equation}
Invoking equation \eqref{duuu}, we immediately see that the partial derivative $\partial \rho/\partial \mathbb{A}^a$, being evaluated at $\mathbb{A}^b=0$, vanishes (recall that $\rho=u/v$). This is a manifestation of the minimum energy principle \cite{Callen_book}. The partial derivative $\partial P/\partial \mathbb{A}^a$ can be rewritten in a more illuminating form. In fact, defined the thermodynamic potential $\mathcal{G}=u+\mathbb{A}^a Y_a$, we have the differential $d\mathcal{G}=Td \mathfrak{s}-Pdv+Y_a d\mathbb{A}^a$, which can be used to derive the following Maxwell relation:
\begin{equation}
\dfrac{\partial P}{\partial \mathbb{A}^a} = -\dfrac{\partial Y_a}{\partial v} \, .
\end{equation}
Physically, this equation is telling us that the susceptibility of the pressure to chemical imbalances equals the susceptibility of the chemical fractions to a volume expansion. Combining these results together, we find that the stress energy tensor in \eqref{tmunu}, expanded to first order in $\mathbb{A}^a$, takes the (Eckart-frame \cite{Eckart40,Bemfica2019_conformal1,
BemficaDNDefinitivo2020}) bulk viscous form
\begin{equation}
T^{\mu \nu} = (\rho_{\text{eq}}+P_{\text{eq}}+\Pi)u^\mu u^\nu +(P_{\text{eq}}+\Pi)g^{\mu \nu} \, ,
\end{equation}
where we are adopting the notation $f_{\text{eq}}(\mathfrak{s},v)= f(\mathfrak{s},v,0)$ and we have introduced the bulk viscous stress
\begin{equation}\label{PiI}
\Pi = -\dfrac{\partial Y_a}{\partial v} \mathbb{A}^a \, .
\end{equation}
Let us now verify explicitly that the mixture indeed admits a Navier-Stokes regime where $\Pi$ is given by \eqref{Pippo}. In order to do this, first we use the chain rule to rewrite equation \eqref{uY} as follows:
\begin{equation}\label{Yaaa}
\dfrac{\partial Y_a}{\partial v} u^\mu \nabla_\mu v + \dfrac{\partial Y_a}{\partial \mathfrak{s}} u^\mu \nabla_\mu \mathfrak{s} + \dfrac{\partial Y_a}{\partial \mathbb{A}^b} u^\mu \nabla_\mu \mathbb{A}^b = v\Xi_{ab}\mathbb{A}^b \, .
\end{equation}
If we retain only the first order terms in $\mathbb{A}^a$ we have that the contribution proportional to $u^\mu \nabla_\mu \mathfrak{s}$ can be neglected, see equation \eqref{us}. Furthermore, we can use the equation $\nabla_\mu n^\mu=0$ to prove that $u^\mu \nabla_\mu v=v\nabla_\mu u^\mu$ \cite{MTW_book}, so that, contracting both sides of \eqref{Yaaa} with $n\Xi^{ab}$ (which is the matrix inverse of $v\Xi_{ab}$), we obtain 
\begin{equation}\label{fourteen}
\begin{split}
& \tau\indices{^a_b} u^\mu \nabla_\mu \mathbb{A}^b + \mathbb{A}^a = \kappa^a \nabla_\mu u^\mu \, ,\\
& \text{with} \quad \tau\indices{^a_b} =-n\Xi^{ac}\dfrac{\partial Y_c}{\partial \mathbb{A}^b} \, , \quad \kappa^a=\Xi^{ab}\dfrac{\partial Y_b}{\partial v} \, . \\
\end{split}
\end{equation}
By the relaxation effect \cite{LindblomRelaxation1996}, we know that in the limit of a very slow process the quantities $u^\mu \nabla_\mu \mathbb{A}^b$ are negligible compared to $\mathbb{A}^a$, so that $\mathbb{A}^a \approx k^a\nabla_\mu u^\mu$. It follows that equation \eqref{PiI} can be approximated as $\Pi \approx -\zeta \nabla_\mu u^\mu$, with
\begin{equation}\label{zetuzza}
\zeta = \Xi^{ab} \dfrac{\partial Y_a}{\partial v}\dfrac{\partial Y_b}{\partial v} \, .
\end{equation}
Clearly, $\zeta$ is non-negative, because $\Xi_{ab}$ (and thus also $\Xi^{ab}$) is positive definite. This completes our proof that a ``slow'' fluid mixture obeys the Navier-Stokes constitutive relations for bulk viscosity.

\section{Burgers-type viscous dynamics}

In the case in which there is only one non-equilibrium fraction $Y_1$, it can be proved that the bulk stress $\Pi$ obeys the Israel-Stewart field equation near local equilibrium \cite{BulkGavassino,GavassinoUniversalityII2023,GavassinoFarFromBulk2023}. Now we will show that, in a similar manner, when there are two fractions $Y_1$ and $Y_2$, the near-equilibrium dynamics of the mixture reproduces the Burgers model.

\subsection{Linearized dynamics about incompressible flows}


In what follows, we will restrict our attention to flows that are ``almost incompressible''.  This means that, fixed a   reference incompressible flow (i.e. fixed an arbitrary solution of the fluid equations with $\nabla_\mu u^\mu=0$), we will study neighbouring compressible solutions to first order in perturbation theory around such reference incompressible flow. This is needed because the Burgers equation is a linear rheological model, and it holds only for small $\nabla_\mu u^\mu$ and $\Pi$.

Let us set up the perturbative expansion rigorously. We consider a smooth one-parameter family of solutions $\{\varphi_i (\epsilon),g_{\mu \nu}(\epsilon)\}$ of the fluid equations coupled with gravity, where $\{\varphi_i(0),g_{\mu \nu}(0)\}$ is an incompressible flow, i.e. $\nabla_\mu u^\mu(0)=0$ across all spacetime.  Such incompressible flow $\{\varphi_i(0),g_{\mu \nu}(0)\}$ may be both fast rotating and accelerating, and it may admit strong shear flows (and large gradients in general), but it does not expand. Note that we are not assuming that the \textit{fluid} itself is incompressible: We are just considering a particular incompressible \textit{solution} of the fluid equations (e.g. a star in hydrostatic equilibrium).  Furthermore, we also assume that $\mathbb{A}^a(0)$ vanishes on some initial Cauchy surface. Then, equation \eqref{fourteen} implies that $\mathbb{A}^a(0)$ vanishes everywhere, meaning that the solution $\epsilon=0$ is reversible: $[u^\mu\nabla_\mu \mathfrak{s}](0)=0$, and also $\Pi(0)=0$, see equation \eqref{PiI}. Thus, since by chain rule
\begin{equation}
u^\mu \nabla_\mu f= \dfrac{\partial f}{\partial v} v \nabla_\mu u^\mu + \dfrac{\partial f}{\partial \mathfrak{s}} u^\mu \nabla_\mu \mathfrak{s} + \dfrac{\partial f}{\partial \mathbb{A}^b} u^\mu \nabla_\mu \mathbb{A}^b \, ,
\end{equation}
we see that all thermodynamic quantities $f(\mathfrak{s},v,\mathbb{A}^b)$ are conserved along the flow worldlines: $[u^\mu \nabla_\mu f](0)=0$. 

Now, we linearise equation \eqref{fourteen} to first order in $\epsilon$, i.e. we differentiate \eqref{fourteen} in $\epsilon$ and evaluate the result at $\epsilon=0$ \cite{Geroch_Lindblom_1991_causal}. This corresponds to studying \eqref{fourteen} in a regime of small compression\footnote{ If the incompressible solution $\{\varphi_i(0),g_{\mu \nu}(0)\}$ is stable, then linear-order perturbation theory is applicable at all times provided that it is applicable on an initial Cauchy surface. If, instead, $\{\varphi_i(0),g_{\mu \nu}(0)\}$ is unstable, then perturbation theory applies only for a finite amount of time (which depends on the Lyapunov exponent of the state), and the Burgers model may break down at late times.}. Introducing the compact notation $Q:=Q(0)$ and $\delta Q:= dQ(0)/d\epsilon$, for any field $Q$, we have the following linear dynamics:
\begin{equation}\label{together}
\tau\indices{^a_b} u^\mu \nabla_\mu \delta \mathbb{A}^b + \delta \mathbb{A}^a =\kappa^a \, \delta(\nabla_\mu  u^\mu) \, .
\end{equation}
Here, $\tau\indices{^a _b}$ and $\kappa^a$ play the role of background quantities, as they are evaluated on the reference incompressible flow (so that $u^\mu \nabla_\mu \tau\indices{^a _b} =u^\mu \nabla_\mu \kappa^a=0$), while $\delta \mathbb{A}^a$ and $\delta(\nabla_\mu  u^\mu)$ are first-order perturbation fields.
In the case of only two non-equilibrium chemical fractions, equation \eqref{together} can be expanded into the system
\begin{equation}
\begin{split}
\tau\indices{^1 _1} u^\mu \nabla_\mu \delta \mathbb{A}^1+\tau\indices{^1 _2} u^\mu \nabla_\mu \delta \mathbb{A}^2+ \delta \mathbb{A}^1 ={}& \kappa^1 \, \delta(\nabla_\mu  u^\mu) \, , \\
\tau\indices{^2 _1} u^\mu \nabla_\mu \delta \mathbb{A}^1+\tau\indices{^2 _2} u^\mu \nabla_\mu \delta \mathbb{A}^2+ \delta \mathbb{A}^2 ={}& \kappa^2  \, \delta(\nabla_\mu  u^\mu) \, .\\ 
\end{split}
\end{equation}
With some simple algebra, we can rewrite the equations above as follows:
\begin{equation}
\begin{split}
(\det \tau) \, u^\nu \nabla_\nu (u^\mu \nabla_\mu \delta\mathbb{A}^1) + (\text{Tr} \, \tau) \, u^\mu \nabla_\mu \delta \mathbb{A}^1 + \delta \mathbb{A}^1 ={}& \kappa^1 \delta(\nabla_\mu  u^\mu) +(\tau\indices{^2 _2 }\kappa^1-\tau\indices{^1 _2}\kappa^2)\, u^\nu \nabla_\nu  \, \delta(\nabla_\mu  u^\mu)\, , \\
(\det \tau) \, u^\nu \nabla_\nu (u^\mu \nabla_\mu \delta\mathbb{A}^2) + (\text{Tr} \, \tau) \, u^\mu \nabla_\mu \delta \mathbb{A}^2 + \delta \mathbb{A}^2 ={}& \kappa^2 \delta(\nabla_\mu u^\mu) +(\tau\indices{^1 _1 }\kappa^2-\tau\indices{^2 _1}\kappa^1)\, u^\nu \nabla_\nu  \, \delta(\nabla_\mu  u^\mu)\, , \\
\end{split}
\end{equation}
where $\det \tau$ and $\text{Tr} \, \tau$ are respectively the determinant and the trace of the matrix $\tau=[\tau\indices{^a _b}]$. Recalling equation \eqref{PiI}, we can finally combine the two field equations for $\delta \mathbb{A}_1$ and $\delta \mathbb{A}_2$ to have an equation of motion for $\delta \Pi$ of the form
\begin{equation}\label{burgers}
\lambda_2 \, u^\nu \nabla_\nu (u^\mu \nabla_\mu \delta\Pi) + \lambda_1 \, u^\mu \nabla_\mu \delta \Pi+ \delta \Pi = -\zeta \, \delta(\nabla_\mu  u^\mu) -\chi\, u^\nu \nabla_\nu  \, \delta(\nabla_\mu  u^\mu)\, ,
\end{equation}
where $\zeta$ is given by equation \eqref{zetuzza}, and
\begin{equation}\label{kalibo}
\begin{split}
\lambda_2 ={}& \det \tau \, ,\\
\lambda_1 ={}& \text{Tr} \, \tau \, ,\\
\chi ={}& (\det \tau) \dfrac{\partial Y_a}{\partial v} (\tau^{-1})\indices{^a _b} \kappa^b  \, . \\
\end{split}
\end{equation}
Equation \eqref{burgers} is the central formula of this manuscript. It tells us that, for a mixture with two non-conserved fractions, the bulk viscous stress $\Pi$ obeys an equation of motion that is of second order in time. This reflects the fact that there are two algebraic non-equilibrium degrees of freedom ($\mathbb{A}^1$ and $\mathbb{A}^2$). Indeed, if one wants to solve equation \eqref{burgers}, they need to prescribe not only the initial value of $\Pi$, but also its initial time derivative, namely $u^\mu \nabla_\mu \Pi$. 

We would like to stress that equation \eqref{burgers} is \textit{not} an expansion in powers of ``$\, u^\mu \nabla_\mu \,$''. In fact, in the derivation, no assumption was made about how fast the process is. The only approximation that we made was the linearization in $\delta \mathbb{A}^a$ and $\delta (\nabla_\mu u^\mu)$, which is an assumption on the amplitude of the perturbation, and not on its frequency. Indeed, equation \eqref{burgers} well approximates \eqref{uY} at all frequencies. Despite this, it is clear that the Burgers model remains applicable also in dynamical regimes with small frequency and large amplitude, as it reduces to Navier-Stokes by the relaxation effect \cite{LindblomRelaxation1996}, with the correct bulk viscosity coefficient $\zeta$.

\subsection{Recovering the Burgers Green function}

Now we only need to show that equation \eqref{burgers} is indeed the Burgers equation for viscoelastic matter. In order to do so, we must prove that its linear-response Green function is \eqref{GtB}. The proof goes as follows. 
The matrices $-n\partial Y /\partial \mathbb{A}= [-n\partial Y_a/\partial \mathbb{A}^b]$ and $\Xi=[\Xi_{ab}]$ are both symmetric and positive definite. Therefore, there exist an invertible real matrix $\mathcal{N}=[\mathcal{N}\indices{_{(a)} ^b}]$ and a diagonal positive definite matrix $\Lambda=\text{diag}\{\tau_{(1)},\tau_{(2)}\}$ such that \cite{horn_johnson_1985}
\begin{equation}
{-}n \dfrac{\partial Y}{\partial \mathbb{A}} = \mathcal{N}^{-1}\mathcal{N}^{-T} \, , \spc \Xi = \mathcal{N}^{-1} \Lambda^{-1} \,\mathcal{N}^{-T} \, .
\end{equation}
Then, the matrix $\Xi^{-1}=[\Xi^{ab}]$ decomposes into $\Xi^{-1}=\mathcal{N}^{T} \Lambda\mathcal{N}$, and we find
\begin{equation}
\tau = \Xi^{-1} \bigg({-}n \dfrac{\partial Y}{\partial \mathbb{A}} \bigg) = \mathcal{N}^T \Lambda \mathcal{N}^{-T} \, .
\end{equation}
Introducing the notation
\begin{equation}
\dfrac{\zeta_{(1)}}{\tau_{(1)}} = \bigg( \mathcal{N}\indices{_{(1)} ^b} \dfrac{\partial Y_b}{\partial v} \bigg)^2 \, , \spc \dfrac{\zeta_{(2)}}{\tau_{(2)}} = \bigg( \mathcal{N}\indices{_{(2)} ^b} \dfrac{\partial Y_b}{\partial v} \bigg)^2 \, , 
\end{equation}
equations \eqref{zetuzza} and \eqref{kalibo} can be rewritten as follows:
\begin{equation}\label{xemo}
\begin{split}
\lambda_2 ={}& \tau_{(1)}\tau_{(2)} \, , \\
\lambda_1 ={}& \tau_{(1)}+\tau_{(2)} \, , \\
\zeta ={}& \zeta_{(1)}+\zeta_{(2)} \, , \\
\chi ={}& \zeta_{(1)}\tau_{(2)}+\zeta_{(2)}\tau_{(1)} \, . \\
\end{split}
\end{equation}
Thus, if we work in a global coordinate system such that $u^\mu=\delta^\mu_t$ on all events, equation \eqref{burgers} reduces to
\begin{equation}\label{zub}
\tau_{(1)}\tau_{(2)}\partial^2_t \delta \Pi +(\tau_{(1)}+\tau_{(2)})\partial_t \delta \Pi +\delta \Pi = -(\zeta_{(1)}+\zeta_{(2)}) \, \delta(\nabla_\mu  u^\mu)-(\zeta_{(1)}\tau_{(2)}+\zeta_{(2)}\tau_{(1)})\partial_t  \, \delta(\nabla_\mu  u^\mu) \, .
\end{equation}
Let us now focus on the Green function \eqref{GtB}. If we plug this choice of $G(t')$ into equation \eqref{babibbo}, we find that $\delta\Pi$ can be expressed as the sum of two contributions, $\delta\Pi_{(1)}$ and $\delta\Pi_{(2)}$, each of which obeys an independent Israel-Stewart-type equation:
\begin{equation}
\begin{split}
(\tau_{(1)}\partial_t +1)\delta\Pi_{(1)}={}& -\zeta_{(1)} \, \delta(\nabla_\mu  u^\mu) \, ,\\
(\tau_{(2)}\partial_t +1)\delta\Pi_{(2)}={}& -\zeta_{(2)} \, \delta(\nabla_\mu  u^\mu) \, .\\
\end{split}
\end{equation}
Applying the operator $\tau_{(2)}\partial_t+1$ to the first equation, and the operator $\tau_{(1)}\partial_t+1$ to the second equation, and adding together the resulting formulas, we indeed recover \eqref{zub}. This shows that the dynamics described by \eqref{burgers} arises from the Burgers Green function \eqref{GtB}, which is what we wanted to prove.

\subsection{Overdamped oscillations}

If we set the right-hand side of \eqref{zub} to zero (incompressible evolution), we find that the bulk stress obeys the equation $\lambda_2 \partial^2_t \delta \Pi + \lambda_1 \partial_t \delta \Pi+\delta \Pi=0$, which describes the dynamics of a damped harmonic oscillator \cite{Denicol_Relaxation_2011}. Such oscillator is necessarily overdamped, i.e. $\lambda_1^2-4\lambda_2 \geq 0$, which follows from equation \eqref{xemo}, and from the fact that $\tau_{(1)}$ and $\tau_{(2)}$ are always positive. The implication is that the bulk stress cannot really ``oscillate''. Instead, the evolution of $\delta \Pi$ is the superposition of two exponential relaxations. Indeed, it is straightforward to show that the chemical mixture has two non-hydrodynamic modes with purely imaginary frequency gap: $\omega_{(a)}(k=0)=-i/\tau_{(a)}$. This is not a surprise, since it is well known that chemical oscillations are forbidden in the linear regime \cite{Yongefeng2008}. This is a consequence of the Onsager symmetry of $\Xi_{ab}$, which forces all non-hydrodynamic gaps to lay on the imaginary axis \cite{GavassinoNonHydro2022}.

\section{Application to neutron-star matter}

Let us now discuss an interesting astrophysical application: bulk viscosity in neutron stars. It is well known that, if neutrinos are not trapped, neutron star matter can be viewed as a fluid mixture of the kind discussed in section \ref{sec2} \cite{Camelio2022,CamelioSimulations2022}, where the conserved current is the baryon current, and the non-equilibrium fractions are the electron fraction $Y_e$ and the muon fraction $Y_\mu$. The corresponding chemical affinities are the $\beta$-reaction affinities, i.e. $\mathbb{A}^e=\mu^n-\mu^p-\mu^e$ and $\mathbb{A}^\mu=\mu^n-\mu^p-\mu^\mu$, see Appendix \ref{AAAAAAAAAAAA} for a quick derivation.

It can be verified (both analytically \cite{BulkGavassino,GavassinoFarFromBulk2023} and numerically \cite{CamelioSimulations2022}) that, if one neglects all muon contributions, the near equilibrium dynamics of $\Pi$ is accurately described by the Israel-Stewart theory\footnote{The fact that escaping neutrinos take energy away does not affect the outcome of the mathematical analysis we carried out till this point. One only needs to correct the conservation law $\nabla_\mu T^{\mu \nu}=0$ with a luminosity term \cite{Camelio2022}: $\nabla_\mu T^{\mu \nu}=-\mathcal{Q}u^\nu$.}. This is indeed expected, since there is only one independent non-equilibrium fraction ($Y_e$), and the fluid falls into the Israel-Stewart universality class \cite{GavassinoUniversalityII2023}.  However, it has also been verified numerically \cite{CamelioSimulations2022} that the Israel-Stewart approximation breaks down completely (also close to equilibrium) if muon contributions are taken into account. Now we are in the position to show that the correct viscous model for neutron-proton-electron-muon ($npe\mu$) matter indeed is not the Israel-Stewart theory, but the Burgers model, i.e. equation \eqref{burgers}\footnote{With analogous calculations to those presented here, it is straightforward to show that the Burgers approximation holds also in the case
of trapped neutrinos, provided that the matter is in the slow lepton-equilibration limit \cite{AlfordMuons2021,AlfordMuons2022}}. 

The analysis that follows formally applies only to hot (i.e. non-superfluid) neutron star matter, e.g. in proto-neutron stars and neutron-star mergers \cite{AlfordMuons2021}. However, superfluidity is not expected to change the hydrodynamic behaviour of the bulk stress $\Pi$ qualitatively \cite{Gusakov2007,GavassinoKhalatnikov2022}, meaning that also superfluid $npe\mu$ matter should behave in a similar way (at least for what concerns bulk viscosity).

\subsection{Dictionary with nuclear physics}

Our goal is to prove that equation \eqref{burgers} exactly reproduces the results of Alford, Harutyunyan, and Sedrakian \cite{AlfordMuons2021,AlfordMuons2022} about bulk viscosity in hot and dense $npe\mu$ matter. To this end, we need first to express the Burgers transport coefficients $\lambda_2$, $\lambda_1$, $\zeta$, and $\chi$ in terms of the quantities $A_n$, $A_p$, $A_1$, $A_2$, $C_1$, $C_2$, $\lambda_e$, and $\lambda_\mu$ introduced in \cite{AlfordMuons2021,AlfordMuons2022}. Comparing our formalism with that of \cite{AlfordMuons2022}, going through some simple (albeit tedious) algebra, one can show that, in the ordered chemical basis $\{e,\mu\}$, the following ``dictionary relations'' hold:
\begin{equation}
\begin{split}
-n\dfrac{\partial Y}{\partial \mathbb{A}} ={}& 
\begin{bmatrix}
A_1 & A_n+A_p \\
A_n+A_p & A_2
\end{bmatrix}^{-1}, \\
-\dfrac{\partial Y}{\partial v} ={}& 
\begin{bmatrix}
A_1 & A_n+A_p \\
A_n+A_p & A_2
\end{bmatrix}^{-1}
\begin{bmatrix}
C_1 \\
C_2 \\
\end{bmatrix} ,\\
\Xi={}&
\begin{bmatrix}
\lambda_e & 0 \\
0 & \lambda_\mu \\
\end{bmatrix} . \\
\end{split}
\end{equation}
Plugging these formulas into \eqref{zetuzza} and \eqref{kalibo}, and introducing the compact notation $\mathfrak{D}=A_1 A_2-(A_n+A_p)^2$, we obtain
\begin{equation}\label{transpero}
\begin{split}
\lambda_2 ={}& (\mathfrak{D} \lambda_e \lambda_\mu )^{-1} \, , \\
\lambda_1 ={}& \dfrac{1}{\mathfrak{D}} \bigg( \dfrac{A_2}{\lambda_e} + \dfrac{A_1}{\lambda_\mu} \bigg) \, , \\
\zeta ={}& \dfrac{[C_1 A_2-C_2(A_n+A_p)]^2}{\mathfrak{D}^2\lambda_e} + \dfrac{[C_2 A_1-C_1(A_n+A_p)]^2}{\mathfrak{D}^2\lambda_\mu} \, , \\
\chi ={}& \dfrac{A_2 C_1^2+A_1 C_2^2-2(A_n+A_p)C_1C_2}{\mathfrak{D}^2 \lambda_e \lambda_\mu} \, . \\
\end{split}
\end{equation}
It is immediate to verify that our formula for $\zeta$ coincides with equation (43) of \cite{AlfordMuons2022}. This  confirms that the Burgers model has the correct ``infrared behaviour'' in the Navier-Stokes limit.

\subsection{Comparison of the effective viscosity coefficients}

In order to compare the dynamics of the Burgers model with that of the multicomponent model of \cite{AlfordMuons2021,AlfordMuons2022}, we need to work in the same physical setting, and compare analogous quantities. The analysis of \cite{AlfordMuons2021,AlfordMuons2022} focuses on small periodic oscillations, so that we need to set $\delta Q \propto e^{-i\omega t}$ (note the different sign convention for $\omega$ in \cite{AlfordMuons2021,AlfordMuons2022}) for all quantities $\delta Q$, where the frequency $\omega \in \mathbb{R}$ is not necessarily small, in the sense that we may also have $|\omega\tau\indices{^a _b}| \gg 1$. Then, equation \eqref{burgers} becomes
\begin{equation}\label{transpasso}
\delta \Pi = - \dfrac{\zeta - i\chi \omega}{1-i\lambda_1\omega -\lambda_2 \omega^2 } \delta(\nabla_\mu u^\mu) \, .
\end{equation} 
If we plug \eqref{transpero} into \eqref{transpasso}, we obtain a complicated formula for $\delta \Pi$, which is \textit{different} from the corresponding equation (37) of \cite{AlfordMuons2022}. The reason is that, in section \ref{dinamo2}, we have defined $\Pi$ as the deviation of the total pressure from the state of chemical equilibrium, while Alford, Harutyunyan, and Sedrakian quantify $\Pi$ as its deviation from the pressure at frozen fractions (see also \cite{AlfordFirst2020}). Hence, we are just comparing different physical quantities. However, if our analysis is correct, the effective viscosity coefficient $\zeta_{\text{eff}}(\omega)$, defined by the relation $\zeta_{\text{eff}} = -\mathfrak{Re} [\delta \Pi/  \delta (\nabla_\mu u^\mu)]$, must be the same, since its value can be inferred from the free-energy dissipation rate \cite{BaggioliHolography,Sawyer_Bulk1989,BulkGavassino}. Indeed, from equation \eqref{transpasso}, we get 
\begin{equation}\label{fedele}
\zeta_{\text{eff}}(\omega)= \dfrac{\lambda_e \lambda_\mu \big\{\lambda_e [(A_n+A_p)C_1-A_1 C_2]^2+\lambda_\mu[(A_n+A_p)C_2-A_2 C_1]^2 \big\}+\omega^2 (\lambda_e C_1^2+\lambda_\mu C_2^2) }{\big\{ \lambda_e \lambda_\mu [A_1A_2-(A_n +A_p)^2]-\omega^2 \big\}^2 +\omega^2(\lambda_e A_1+\lambda_\mu A_2)^2} \, ,
\end{equation}
which perfectly agrees with equation (40) of \cite{AlfordMuons2022}. This shows that $npe\mu$ matter is accurately described, at all hydrodynamic frequencies, by the Burgers model. Needless to say that, instead, the Israel-Stewart theory, whose effective viscosity coefficient is a Lorentzian function, $\zeta_{\text{eff}}(\omega)=\zeta/(1+\omega^2 \tau^2)$ \cite{BulkGavassino}, cannot reproduce equation \eqref{fedele}, unless $\lambda_e$ or $\lambda_\mu$ vanishes or diverges.  

The astrophysical example of $npe\mu$ matter discussed above shows that bulk viscous fluids of Burgers type are not a mere mathematical conjecture. They exist in nature. More importantly, deviations from both Navier-Stokes and Israel-Stewart enter the formula for $\zeta_{\text{eff}}(\omega)$, thereby modifying the dissipation rate of sound waves. Whether these effects have a measurable impact on the damping rate of neutron star oscillations is left as a future direction of investigation.

\section{Conclusions}

All fluids are non-Newtonian. The question is whether the hydrodynamic process under consideration explores frequencies for which a gradient expansion like \eqref{Pippo} is applicable. When this is not the case, conventional relativistic hydrodynamics (as it is formulated e.g. in \cite{Romatschke2017}) ceases to exist, and we enter the domain of rheology. In a rheological context, there is no hope for a universal theory applicable to all fluids. Instead, one deals with a population of different universality classes \cite{GavassinoUniversalityI2023} (called ``rheological models''  \cite{Malkin_book}), of which the Israel-Stewart theory \cite{Israel_Stewart_1979} is a famous example.

In neutron stars, the relaxation time associated to $\beta$-processes is slow \cite{Sawyer_Bulk1989} (being governed by the weak interaction) and it may become comparable to the timescale of the hydrodynamic processes occurring, e.g., in a merger \cite{AlfordRezzolla,MostHaber2022}. Hence, we need rheology. We have proved that, if one accounts for the presence of muons, the rheological model that properly describes the dynamics of the bulk stress in neutron-star matter near local equilibrium is not the Israel-Stewart theory (or Maxwell model), but the Burgers model for viscoelasticity:
\begin{equation}\label{lamb2}
\lambda_2 \, u^\nu \nabla_\nu (u^\mu \nabla_\mu \Pi) + \lambda_1 \, u^\mu \nabla_\mu \Pi+  \Pi = -\zeta  \nabla_\mu  u^\mu -\chi\, u^\nu \nabla_\nu  (\nabla_\mu  u^\mu) \, .
\end{equation}
The four transport coefficients $\lambda_2$, $\lambda_1$, $\zeta$, and $\chi$ can be expressed in terms of nuclear reaction rates and susceptibilities through equation \eqref{transpero}, where $A_n$, $A_p$, $A_1$, $A_2$, $C_1$, $C_2$, $\lambda_e$, and $\lambda_\mu$ are computed in \cite{AlfordMuons2021,AlfordMuons2022}. More in general, we have shown that any reactive mixture having one single conserved current and two non-conserved independent fractions can be reinterpreted (near local equilibrium) as a bulk viscous fluid of Burgers type. The  formulas for the transport coefficients, expressed in terms of the chemical kinetic coefficients, are reported in equations \eqref{zetuzza} and \eqref{kalibo}.

Viscous fluids of Burgers type are particularly interesting from a fluid-dynamical perspective because they exhibit qualitative behaviours that are strictly forbidden within Israel-Stewart phenomenology. In particular:
\begin{figure}
\begin{center}
	\includegraphics[width=0.49\textwidth]{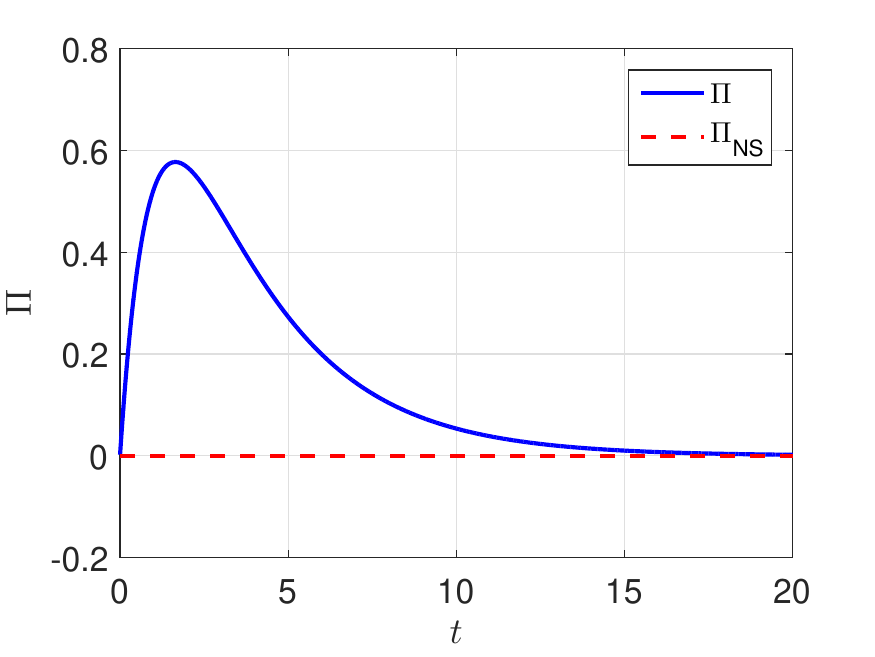}
		\includegraphics[width=0.49\textwidth]{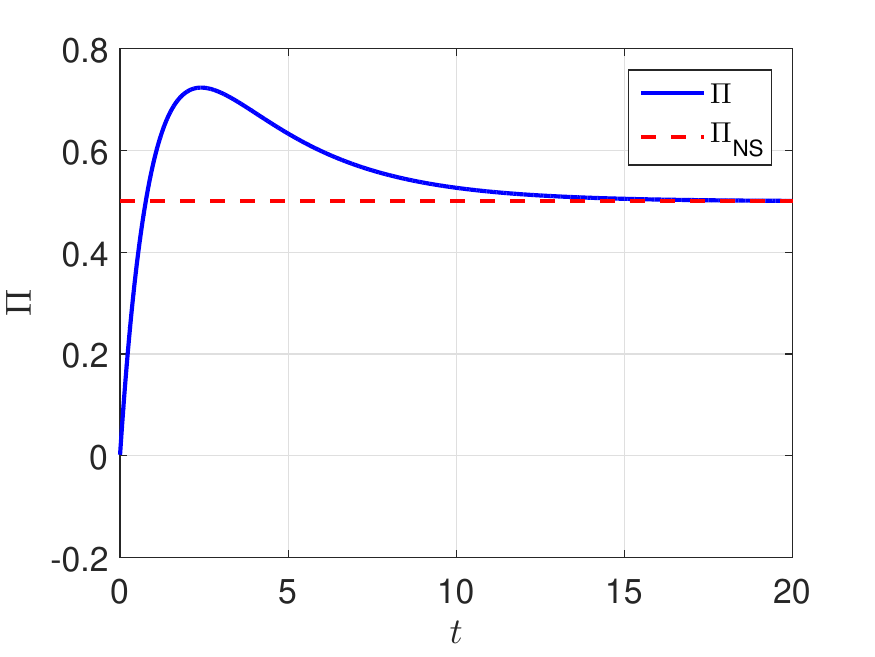}
	\caption{Two solutions of the linearised Burgers equation $3\ddot{\Pi}(t)+4\dot{\Pi}(t)+\Pi(t)=\Pi_{NS}=\text{const}$ whose qualitative behaviour is forbidden within Israel-Stewart hydrodynamics. Left panel: An incompressible flow ($\Pi_{NS}=0$) where the viscous stress spontaneously departs from zero. This can happen because the Burgers equation is of second order, so that, even if $\Pi(0)=0$, we can set $\dot{\Pi}(0)=1$. Right panel: Stress overshoot. The bulk stress does not relax directly to $\Pi_{NS}=1/2$, but it temporarily evolves above it. This happens if the initial value of $\dot{\Pi}$ is larger than $1/2$, in which case the bulk stress reaches $\Pi_{NS}$ ``too fast''.}
	\label{fig:wow}
	\end{center}
\end{figure}
\begin{itemize}
\item[(i)] A bulk viscous fluid of Burgers type can be out of local thermodynamic equilibrium also when $\Pi=0$. From a ``chemistry perspective'', this is obvious. It happens whenever
\begin{equation}
0 \neq \mathbb{A}^2= -\dfrac{\partial Y_1/\partial v}{\partial Y_2/\partial v} \mathbb{A}^1 \, ,
\end{equation}  
see equation \eqref{PiI}. However, from the perspective of an effective viscous description, this is highly non-trivial, because it implies that the stress $\Pi$ can spontaneously depart from zero even during an incompressible flow, and in the absence of external agents (see figure \ref{fig:wow}, left panel). Another consequence is that the entropy production rate in Burgers materials is not proportional to $\Pi^2$ (even close to equilibrium), because irreversible dynamics can occur also at zero $\Pi$. 
 \item[(ii)] Since equation \eqref{lamb2} is of second order in time, the assumption of Extended Irreversible Thermodynamics \cite{Zakari1993,Jou_Extended,rezzolla_book} that the bulk stress $\Pi$ just ``relaxes'' towards its Navier-Stokes value, $\Pi_{NS}=-\zeta \nabla_\mu u^\mu$, is violated by Burgers materials. For example, we can have a situation where $\Pi$ overshoots $\Pi_{NS}$ (see figure \ref{fig:wow}, right panel). Indeed, the possibility of a stress overshoot in viscoelastic materials has been observed experimentally \cite{Falk2011,Divoux2011}, and it is predicted by many holographic models \cite{BaggioliHolography}. The analysis we carried out here sheds new light on this phenomenon, as it reveals that stress overshoots occur in all those thermodynamic systems that have (at least) two non-equilibrium degrees of freedom which relax on timescales $\tau_{(1)}\neq \tau_{(2)}$ of the same order of magnitude.
\end{itemize}
We would like to stress that, for linear deviations about equilibrium, the field equations of the Burgers model are \textit{mathematically} equivalent to the field equations of the chemical mixture. This implies that, if the equation of state of the mixture is thermodynamically consistent \cite{landau5,GavassinoLyapunov_2020,
GavassinoStabilityCarter2022,
GavassinoGENERIC2022}, then the corresponding Burgers model is necessarily causal \cite{GavassinoCausality2021}, covariantly stable \cite{GavassinoGibbs2021,GavassinoSuperluminal2021,GavassinoBounds2023}, and symmetric hyperbolic \cite{GavassinoCasmir2022}, at least in the linear regime.

At this point, one may wonder whether it is necessary that we implement the Burgers model into numerical codes which describe neutron star oscillations and mergers. In practice, this is not needed. In fact, the Burgers model is a near-equilibrium approximation of the chemical evolution of $npe\mu$ matter (see section \ref{dynamo}). Numerical codes that explicitly track the chemical fractions $Y_a$ already exist \cite{Perego19,RadiceEjecta2,MostHaber2022}, and they will automatically reproduce the detailed Burgers dynamics in the appropriate regimes. Indeed, our analysis shows that, even close to local equilibrium, effective viscous theories such as Israel-Stewart \cite{Hishcock1983} or BDNK \cite{BemficaDNDefinitivo2020} (whose bulk sector is equivalent to that of Israel-Stewart \cite{DoreGavassino2022}) may not be reliable, and it is always ``safer'' just to track all chemical components explicitly. On the other hand, we believe that the Burgers model can be a handy tool in analytical models of  neutron star oscillations, since it is easier to implement the backreaction of chemistry onto the flow directly as an effective viscous sector.

\section*{Acknowledgements}

This work was supported by a Vanderbilt's Seeding Success Grant. I would like to thank S. Harris, M. Alford and A. Harutyunyan for fruitful exchanges. I am also grateful to M. Disconzi, B. Haskell, and M. Antonelli for reading the manuscript and providing useful comments.

\appendix

%

\section{Affinities of neutron star matter}\label{AAAAAAAAAAAA}

Since the energy of escaping neutrinos is not accounted for, the differential of the energy density of $npe\mu$ matter is 
\begin{equation}\label{npemu}
d\rho=Tds+\mu^n dn_n+\mu^pdn_p+\mu^e dn_e+\mu^\mu dn_\mu \, ,
\end{equation}
where $T$ is the temperature, $s$ is the entropy density, $\mu^i$ and $n_i$ are chemical potentials and densities of neutrons ($n$), protons ($p$), electrons ($e$), and muons ($\mu$). If we enforce charge neutrality, i.e. $n_p=n_e+n_\mu$, and if the only conserved density is the baryon number $n=n_n+n_p$ (because neutrinos can carry away lepton number), the differential above can be rewritten as follows:
\begin{equation}
d\rho= Tds+\mu^n dn-\mathbb{A}^edn_e-\mathbb{A}^\mu dn_\mu \, ,
\end{equation}
where $\mathbb{A}^e=\mu^n-\mu^p-\mu^e$ is the affinity of the reactions $p+e \rightarrow n+\nu$ and $n \rightarrow p+e+\bar{\nu}$, while $\mathbb{A}^\mu=\mu^n-\mu^p-\mu^\mu$ is the affinity of the reactions $p+\mu \rightarrow n+\nu$ and $n \rightarrow p+\mu+\bar{\nu}$ \cite{Camelio2022}. If we switch to quantities per baryon, we immediately see that the differential of the specific energy $u$ has the form \eqref{duuu}, i.e.
\begin{equation}
du = Td\mathfrak{s}-Pdv-\mathbb{A}^e dY_e-\mathbb{A}^\mu dY_\mu \, .
\end{equation}

\bibliography{Biblio}

\label{lastpage}

\end{document}